\documentclass[12pt]{article}

\usepackage{eurosym}
\usepackage[totalwidth=460truept,totalheight=600truept]{geometry}
\usepackage{latexsym,graphicx,amsfonts,amssymb,amsmath,xcolor}
\usepackage[hypertexnames=false,hidelinks]{hyperref}
\usepackage{fancyhdr}

\global\arraycolsep=1truept

\begin{document}

\bigskip \phantom{C}

\vskip 2truecm

\begin{center}
{\huge \textbf{On Causality and Predictivity}}

\vskip 1.5truecm

\textsl{Damiano Anselmi}

\vskip .1truecm

\textit{Dipartimento di Fisica \textquotedblleft Enrico
Fermi\textquotedblright , Universit\`{a} di Pisa}

\textit{Largo B. Pontecorvo 3, 56127 Pisa, Italy}

\textit{and INFN, Sezione di Pisa,}

\textit{Largo B. Pontecorvo 3, 56127 Pisa, Italy}

damiano.anselmi@unipi.it

\vskip 3truecm

\textbf{Abstract}
\end{center}

Certain approaches to quantum gravity, such as the one based on the concept
of purely virtual particles (fakeons), sacrifice the cause-effect relation
at very small scales to reconcile renormalizability with unitarity. Other
developments have also urged caution regarding the idea of causality as a
fundamental principle. In this paper, we examine the problem from multiple
perspectives, including locality and predictivity, and extend the existing
skepticism in several directions. Emphasizing the impact of unruly
\textquotedblleft disruptors\textquotedblright , we point out that the
illusory arrow of time associated with causality and predictivity is
inherently statistical. This renders the cause-effect relation strained at
the microscopic level. We also show that causation is a borderline concept
that demands belief in entities which can act on nature without being part
of it. Ultimately, not only is renouncing microcausality a reasonable price
to pay for a consistent and predictive theory of quantum gravity (as is the
one based on the fakeon idea), but the very notion of causality is
misleading. Resting as it does on metaphysical assumptions, it should
therefore be abandoned in fundamental physics.

\vfill\eject

\section{Introduction}

\label{intro}\setcounter{equation}{0}

Among the foundational principles of Quantum Field Theory (QFT) are
locality, renormalizability, and unitarity. While unitarity is a
self-consistency requirement, essential for probability conservation,
locality and renormalizability are more pragmatic principles that have
successfully guided the construction of theories describing three of the
four fundamental interactions of nature. Microcausality, often closely
linked to locality and analyticity, has historically played an important,
though perhaps less examined, role. However, for various reasons, the
necessity of strict causality is increasingly being questioned in
fundamental physics, particularly in relation with quantum gravity (QG).

Although it is generally anticipated that quantum gravity may demand the
sacrifice of certain fundamental principles, this necessity does not
automatically mandate the abandonment of the entire QFT framework, which has
proven so successful in describing the other three interactions. This
position stands in contrast to the assumptions underlying approaches such as
String Theory, Loop Quantum Gravity, and Holography, which necessitate
radical foundational shifts. Such broad departures from QFT frequently lack
predictive power and may reflect an undue pessimism regarding the
flexibility of the standard QFT approach.

It is clear that tackling quantum gravity from within the QFT framework
presents a reduced space of maneuver. On the other hand, it is precisely
this powerful constraint on available options that can lead to a predictive
outcome. Should the necessary theoretical renunciation prove to be minimal,
the approach's effectiveness would be even more convincing.

Among the principles that may be readily relaxed are locality, causality,
and analyticity, provided they are not abandoned entirely but merely
modified to the minimum indispensable extent. While analyticity may be
viewed as partially dispensable, given its status as a pragmatic
requirement, causality and, to some extent, locality, may point to more
fundamental principles that some may not be willing to renounce so quickly,
not even at the microscopic level.

What is a \textquotedblleft cause\textquotedblright ? This seemingly
innocent question conceals more complexity than is customarily assumed.
Since Hume \cite{Hume}, the answer should not be taken for granted. Is
causation a practical shortcut, as Hume claims, or a fundamental principle
of nature? This\ is not a minor problem, as elevating a practical shortcut
to the status of a physical law --- or even a principle --- carries the risk
of misdirection, potentially obscuring promising candidate solutions to
existing open problems.

Nonrelativistic and relativistic mechanics do not prompt significant
suspicion about causality. Quantum mechanics starts to plant a seed of
doubt, because its standard interpretation includes the non-causal
(probabilistic and instantaneous) collapse of the wave function upon
measurement, fundamentally breaking the deterministic evolution governed by
the Schr\"{o}dinger equation. Quantum field theory raises more concerns,
since it lacks a convincing and broadly accepted definition of
microcausality. Despite this conceptual ambiguity, QFT has achieved
remarkable success and predictive power. Given that substantial progress can
be made without resolving the issue, the strict necessity of the causality
concept is, at the very least, redundant for fundamental physics. Quantum
gravity marks a step ahead. In many approaches, strict causality is not
expected to be crucial at or below the Planck length. Some frameworks
predict that causality may be broken at much larger scales.

For example, in quantum gravity with fakeons \cite{LWgrav,Absograv} (i.e.,
purely \ virtual particles, or particles that can never be on the mass shell 
\cite{AP,diagrammarMio,PVP20}), the scale of causality violation is $%
1/m_{\chi }$, where $m_{\chi }$ is the mass of the \textquotedblleft
gravifakeon\textquotedblright , a spin-2 purely virtual particle that
belongs to the fundamental triplet of the theory (graviton, inflaton and
fakeon), necessary to ensure consistency through unitarity and
renormalizability. On cosmological grounds \cite{ABP}, $m_{\chi }$ is
constrained to be larger than $m_{\phi }/4$, where $m_{\phi }$ denotes the
mass of the inflaton (approximately 10$^{13}$GeV according to the results on
primordial spectra of scalar fluctuations \cite{Planck}). Ultimately, the
fakeon approach allows for violations of causality that extend up to six
orders of magnitude above the Planck length.

Another approach for building QG\ models that are both renormalizable and
unitary relies on removing the ghosts of higher-derivative local theories by
inserting appropriate nonpolynomial form factors into the propagators \cite%
{nonlocal}. The resulting theories are nonlocal and violate microcausality,
albeit in different ways.

Nonlocal QFT has been attracting attention for a long time, from the
pioneering works of Pais and Uhlenbeck \cite{Pais} and Efimov \cite{efimovnl}%
. Interest in this area has been revived more recently by many authors \cite%
{nonlocal,nonlocrest}. The main problem with a plain nonlocal approach is
that it entails an infinite degree of arbitrariness, and no physical
principle is currently known to single out a unique theory. This contrasts
sharply with the fakeon framework, which yields a unique strictly
renormalizable QG\ model \cite{LWgrav}. It has been argued \cite{nonlocMio}
that if a nonlocal unitary theory has a regular local limit, that limit must
be a model containing fakeons. Then the local limit provides the missing
criterion for selecting the \textquotedblleft right
theory\textquotedblright\ within the infinite space of nonlocal theories.

Acausal behaviors and related phenomena are encountered in several other
contexts. Among these, we mention the Lee-Wick models \cite{LeeWick}, in
which the \textquotedblleft abnormal particles\textquotedblright\ rapidly\
decay. Further examples are approaches based on propagators with complex
poles \cite{Veltman}, analogies with QCD \cite{Holdom}, antilinear
symmetries \cite{Mannheim} and unstable ghosts \cite{Donoghue}. It is worth
noting that a violation of causality is not expected in the Stelle theory 
\cite{Stelle}, which is quadratic gravity with a spin-2 massive ghost. The
primary trouble with that theory, however, is its lack of unitarity.

In this paper, we critically examine the concept of causality within both
deterministic and non-deterministic frameworks. We argue that the notion of
cause is only tenable when it refers to entities that are genuinely external
to the system under observation. Crucially, these external entities must not
be subject to the laws of physics. Otherwise, they would merely constitute
internal components of larger systems, thereby losing their essence as
causes. Because the universe contains no truly external entity capable of
acting upon the universe itself, we conclude that there are no true causes
within nature.

The idea of cause, therefore, is a borderline concept. First, it belongs to
our description of nature rather than being an intrinsic feature of nature
itself. However, just as the laws of physics are invariant under changes of
coordinates and reference frames, it is evident that nature does not depend
on our ways to formulate it. Second, describing nature in terms of causes
treads into metaphysics, or \textquotedblleft the
supernatural\textquotedblright , since it involves entities that are outside
nature, but can act on it.

The cause-effect relation affords us the illusion of having control over the
future by changing the course of events. Once we accept that this is, in
fact, an illusion, we could settle for simply \textit{predicting} the
future. Motivated by this consideration, we critically examine the issue of
predictivity, and argue that true predictions are also impossible as a
matter of principle. Instead, we can only make statements that can be
verified retrospectively (prepostdictions). This limitation arises because
physical systems can never be perfectly isolated, nor can their initial
conditions be fixed completely. The possibility that unruly
\textquotedblleft disruptors\textquotedblright\ might emerge from regions of
spacetime beyond our control and alter the final outcome in unforeseen ways
can never be rigorously excluded, although it can be practically dismissed
on statistical grounds. This very fact, however, confirms that the illusory
arrow of time attributed to causality is an effect of statistics, emerging
at macroscopic scales. This renders microcausality disposable, even in the
presence of external forces.

In theories with fakeons the maximum we can achieve is delayed
prepostdictions. However, the delay expected in quantum gravity is so short (%
$1/m_{\chi }\lesssim 10^{-37}$ seconds) that it does not significantly
worsen the fundamental limits on our predictive capabilities.

One may object that, for everyday purposes and indeed many established
scientific applications, these are distinctions without a practical impact,
in the sense that we can still understand one another by talking about
causes and effects, and predictions. The problems we discuss here gain
significance when causality is elevated to a fundamental principle. If we
want to decipher quantum gravity, or anyway explore the unknown, what is
sufficient for practical utility may well be insufficient for fundamental
investigation. This caution should not come as a surprise, since quantum
theory has already taught us to be wary of taking anything for granted.

Causality often overlaps with non-superluminality, the property that signals
cannot propagate faster than light in Special Relativity. Yet
non-superluminality is not inherent in the notion of causality. If one
assumes that relations of cause and effect do exist, then
non-superluminality merely restricts them to events lying within the past
and future light cones. On the other hand, if we accept, as we demonstrate
in this paper, that relations of cause and effect lack a fundamental
meaning, we are not thereby forced to accept superluminal propagation.

The arguments advanced in this paper should not be construed as contentious,
as there is little need to convince the majority of the physics community on
the opportunity of relaxing the requirement of microcausality, especially in
quantum gravity. Apart from specific, perhaps dogmatic, segments within
String Theory and Loop Quantum Gravity, which appear to maintain an
a-priori, unexamined commitment to strict causality, skepticism about
causation as a fundamental principle is widespread. Moreover, the broad
community response has been encouragingly receptive and open-minded since
the introduction of the fakeon framework in 2017. The present work,
therefore, aims not at widespread persuasion, but rather to sharpen the
articulation of points that many are already well-disposed to accept,
emphasizing that if the price for achieving a consistent and, crucially,
experimentally testable theory of quantum gravity is to relax the constraint
of strict microcausality, as the fakeon theory requires, it is a trade-off
they are willing to accept. At the same time, we caution upfront that\ we do
not stop there, but push the claim much further, suggesting that the notion
of causality is inherently misleading (insofar as it relies on metaphysical
assumptions) and should therefore be completely abandoned in fundamental
physics.

Among past attempts to formalize physical laws without relying on a
fundamental arrow of causality we mention the Feynman-Wheeler absorber
theory \cite{Wheeler}, which introduces the notion of \textquotedblleft
bidirectional causality\textquotedblright\ by treating advanced and retarded
potentials on an equal footing. In such frameworks, the radiation process is
not a one-way emission but a time-symmetric interaction between an emitter
and an absorber. The problem of nonlocality has also been addressed by
challenging causality in the context of Bell's theorem \cite{Drezet},
exploiting the past-future (anti)symmetry and establishing a relation with
pioneering works by de Broglie \cite{broglie}.

The paper is organized as follows. In Section \ref{det}, we demonstrate that
the notion of cause is untenable in a deterministic theory. In section \ref%
{need}, we show that it hints at the existence of truly external entities,
and point out the conceptual concerns and contradictions implied by this
necessity. Section \ref{prepost} examines the issue of predictivity in
\textquotedblleft causal\textquotedblright\ theories. This discussion is
extended to theories incorporating fakeons in Section \ref{delprepost}. In
Section \ref{cauqtf}, we comment on the difficulties posed by defining
causality in quantum field theory and quantum gravity, while in Section \ref%
{loc} we discuss the relationship with nonlocality. Section \ref{concl}
contains the conclusions.

\section{Determinism and causality}

\setcounter{equation}{0}\label{det}

While aspects of causality were questioned earlier, the Scottish empiricist
David Hume was the first to articulate a rigorous philosophical skepticism
about it. Here are his words, from \textit{An Enquiry Concerning Human
Understanding }\cite{Hume}.

\textquotedblleft All events seem entirely loose and separate. One event
follows another; but we never can observe any tie between them. They seem 
\textit{conjoined}, but never \textit{connected}. And as we can have no idea
of any thing which never appeared to our sense or inward sentiment, the
necessary conclusion \textit{seems} to be that we have no idea of connexion
or power at all, and that these words are absolutely without any meaning,
when employed either in philosophical reasonings or common
life.\textquotedblright\ [Section VII, Part II, p. 76]

\textquotedblleft When we look about us towards external objects, and
consider the operation of causes, we are never able, in a single instance,
to discover any power or necessary connexion; any quality, which binds the
effect to the cause, and renders the one an infallible consequence of the
other. We only find, that the one does actually, in fact, follow the
other.\textquotedblright\ [Section VII, Part I, p. 64]

\textquotedblleft It appears, then, that this idea of necessary connexion
among events arises from a number of similar instances which occur of the
constant conjunction of these events; nor can that idea ever be suggested by
any one of these instances, surveyed in all possible lights and positions.
But there is nothing in a number of instances, different from every single
instance, which is supposed to be exactly similar; except only, that after a
repetition of similar instances, the mind is carried by habit, upon the
appearance of one event, to expect its usual attendant, and to believe that
it will exist. This connexion, therefore, which we \textit{feel} in the
mind, this customary transition of the imagination from one object to its
usual attendant, is the sentiment or impression from which we form the idea
of power or necessary connexion.\textquotedblright\ [Section VII, Part II,
p. 77]

\textquotedblleft Suitably to this experience, therefore, we may define a
cause to be \textit{an object, followed by another, and where all the
objects similar to the first are followed by objects similar to the second}.
Or in other words \textit{where, if the first object had not been, the
second never had existed}. The appearance of a cause always conveys the
mind, by a customary transition, to the idea of the
effect\textquotedblright\ [Section VII, Part II, p. 79]

\textquotedblleft All inferences from experience, therefore, are effects of
custom, not of reasoning. Custom, then, is the great guide of human life. It
is that principle alone, which renders our experience useful to us, and
makes us expect, for the future, a similar train of events with those which
have appeared in the past.\textquotedblright\ [Section V, Part I, pp. 44-45]

\bigskip

In other words, Hume posits that causation is a psychological habit, not an
intrinsic feature of reality. Yet, he seeks to save the idea of cause by
assigning it a role in the \textquotedblleft trains of events in
succession\textquotedblright\ that we encounter through experience. 

The first observation that comes to mind is that this train of events\ more
precisely describes determinism, rather than causality. The distinction
between the two concepts is crucial for our analysis.

In a deterministic world the present is uniquely determined by both the past
and the future. Hence, it is redundant to claim that the present is
\textquotedblleft caused\textquotedblright\ by the past, given that no
alternative was available. Events are merely parts of a fixed chain, and we
interpret them as causes only because of how we experience time. We say, for
example: \textquotedblleft exposing yourself to fresh air today causes you
to be sick tomorrow\textquotedblright . But we could equally well say:
\textquotedblleft you will be sick tomorrow, because it is already written,
and for that reason you exposed yourself to fresh air
today.\textquotedblright\ The direction of the cause-effect relationship is
thus arbitrary, contrary to what causation is supposed to be: a
chronological ordering equipped with an arrow signifying a necessary link
from the past, through the present, to the future. It is our narration of
the universe that equips the flow of events with an apparent arrow.

The fundamental laws of physics (aside from the T violation predicted by the
Standard Model of elementary particles) are symmetric under time reversal.
In classical electrodynamics, standard boundary conditions at infinity force
the use of the retarded potentials over the advanced ones. This choice,
however, has no inherent connection to causation. In other situations, such
as inside a cavity with reflecting walls, or describing how smartphones emit
and receive, we need combinations of both retarded and advanced potentials.
These combinations also have no relation with causality, nor do they imply
its violation. They simply reflect the fact that the electromagnetic field
is mathematically described as a superposition of incoming and outgoing
waves.

For definiteness, consider a pointlike oscillating dipole $\mathbf{d}$
placed at the origin. The \textquotedblleft source\textquotedblright 
\begin{equation}
J^{\mu }(t,\mathbf{r})=-(\cos (\omega t)\mathbf{d}\cdot \mathbf{\nabla }%
\delta ^{(3)}(\mathbf{r}),\omega \mathbf{d}\sin (\omega t)\delta ^{(3)}(%
\mathbf{r})),\qquad \partial _{\mu }J^{\mu }=0,  \label{source}
\end{equation}%
gives the vector potential%
\begin{equation}
A_{-}^{\mu }(t,\mathbf{r})=-\frac{1}{4\pi }\left( \mathbf{\nabla }\cdot 
\frac{\mathbf{d}\cos (\omega (t-r))}{r},\frac{\omega \mathbf{d}\sin (\omega
(t-r))}{r}\right) ,  \label{Am}
\end{equation}%
upon solving $\Box A^{\mu }=J^{\mu }$ in the Lorenz gauge $\partial _{\mu
}A^{\mu }=0$, where $r=|\mathbf{r}|$. However, the same current (\ref{source}%
) also gives the solution\footnote{%
The transformation $t\rightarrow -t$, $\mathbf{r}\rightarrow -\mathbf{r}$, $%
\mathbf{d}\rightarrow -\mathbf{d}$ implies $J^{\mu }\rightarrow J^{\mu }$, $%
\partial _{\mu }\rightarrow -\partial _{\mu }$, $\Box \rightarrow \Box $, $%
A_{\pm }^{\mu }\rightarrow A_{\mp }^{\mu }$.}%
\begin{equation}
A_{+}^{\mu }(t,\mathbf{r})=-\frac{1}{4\pi }\left( \mathbf{\nabla }\cdot 
\frac{\mathbf{d}\cos (\omega (t+r))}{r},\frac{\omega \mathbf{d}\sin (\omega
(t+r))}{r}\right) ,  \label{Ap}
\end{equation}%
in which case it is not a \textquotedblleft source\textquotedblright , but a
\textquotedblleft sink\textquotedblright .

Thus, is (\ref{source}) the \textquotedblleft cause\textquotedblright\ or
the \textquotedblleft end\textquotedblright , the emitter or the receiver?
If (\ref{source}) were the cause, it would suffice to cause the
electromagnetic field encoded in the vector potential (\ref{Am}). Instead,
the choice between (\ref{Am}) and (\ref{Ap}) rests on the boundary
conditions at infinity. Hence, the current (\ref{source}) \textit{per se} is
not a cause.

Remarks along these lines, including the irrelevance of causality for
physics, were articulated more than a century ago by philosophers such as
Bertrand Russell and his followers. Here are some excerpts from Russell's
essay \textquotedblleft On the notion of cause\textquotedblright\ \cite%
{Russell}. 1) \textquotedblleft In the motions of mutually gravitating
bodies, there is nothing that can be called a cause, and nothing that can be
called an effect; there is merely a formula. Certain differential equations
can be found, which hold at every instant for every particle of the system,
and which, given the configuration and velocities at one instant, or the
configurations at two instants, render the configuration at any other
earlier or later instant theoretically calculable. That is to say, the
configuration at any instant is a function of that instant and the
configurations at two given instants. This statement holds throughout
physics, and not only in the special case of gravitation.\textquotedblright\
2) \textquotedblleft All philosophers, of every school, imagine that
causation is one of the fundamental axioms or postulates of science, yet,
oddly enough, in advanced sciences such as gravitational astronomy, the word
`cause' never appears.\textquotedblright\ 3) \textquotedblleft The reason
why physics has ceased to look for causes is that, in fact, there are no
such things. The law of causality, I believe, like much that passes muster
among philosophers, is a relic of a bygone age, surviving, like the
monarchy, only because it is erroneously supposed to do no
harm.\textquotedblright\ 4) \textquotedblleft The word `cause' is so
inextricably bound up with misleading associations as to make its complete
extrusion from the philosophical vocabulary desirable.\textquotedblright

Unfortunately, Russell's arguments failed to open a breach in the
methodology of physics. This failure resulted in the continued adoption of 
\textit{ad hoc} definitions for the sake of preserving the notion of cause,
rather than challenging the concept itself at its core. We believe that it
is about time to settle the matter and eliminate causality from the
foundations of physical sciences.

As noted, the essence of determinism is the lack of alternatives. What if
alternatives were available? Then perhaps one could make sense of concepts
such as cause and effect, because it would be possible to \textquotedblleft
change the course of events\textquotedblright . Nevertheless, the only realm
where multiple outcomes may follow the same initial conditions in physics is
quantum theory. There, the selection among several possibilities is
determined\ by chance, i.e., it occurs without cause. Although the course of
events is not written in advance, it is not possible to control it or guide
it. We must therefore concede that quantum theory is not a promising
candidate to resurrect the notion of causality.

It is worth stressing that time does not possess an arrow in quantum
mechanics either. It is true that the choice of a particular future among
many options (e.g., the choice between left and right in the Stern-Gerlach
experiment) is created \textit{ex nihilo}. At first sight, this uncaused
selection may appear to endow time with a direction. However, this is not
true, because we can mirror the statement by looking backwards in time: the
present may originate from different pasts, and it is impossible to uniquely
trace back the past that led to a particular present.

Consider an electron that is prepared with spin $+1/2$ along the $z$
direction. If we then plan to measure its spin along the $x$ direction, we
cannot predict whether the result will be $+1/2$ or $-1/2$. This
unpredictability is due to the state being a superposition of $x$-spin
eigenstates. Symmetrically, if we find that the spin is, say, $-1/2$ along
the $x$ direction, that knowledge alone does not allow us to uniquely infer
its preceding state: it is impossible to determine whether the particle was
preceded by spin $+1/2$ or $-1/2$ along the $z$ direction. This perfect
symmetry in the uncertainty of determination -- both forward in time
(prediction) and backward in time (retrodiction) -- highlights the acausal
nature of quantum measurement without an inherent arrow of time.

\section{Causes as truly external entities}

\setcounter{equation}{0}\label{need}

Some light is shed on our pondering by noticing that talking about causes
and effects can only make sense if there is a clear distinction between the
system under observation and something \textit{external} to the system, such
as a source or a force. To better explain this point, we consider two
systems that involve modifications to the second Principle of Dynamics, $%
ma=F $, and are supposed to violate causality. One is Dirac's method to
remove runaway solutions in classical electrodynamics \cite{DiracRun,calca}.
The other is the case of fakeons, or purely virtual particles, which stand
as promising tools for a variety of applications.

It is useful to compare these two systems face to face to highlight their
differences and commonalities. While fakeons are claimed to be fundamental
entities, Dirac's theory is an effective description of the
\textquotedblleft friction\textquotedblright\ and energy loss due to the
emission of radiation by an accelerating charged particle. Furthermore, in
Dirac's case time possesses an arrow (a detail which is irrelevant to our
main argument, but which might otherwise inject a source of confusion into
the discussion), whereas fakeons are invariant under\ time reversal. Other
differences (e.g., Dirac's treatment is purely classical, while fakeons,
which are due to unitarity in quantum field theory, do not have a classical
counterpart, yet affect ordinary particles indirectly) are not crucial for
the discussion of this paper:\ both systems are valuable for illustrating
the conceptual points we wish to make.

Let us start from $ma(t)=F(t)$, where $F(t)$ is an external force. We may
assert that $F$ is the cause and the acceleration $a=\ddot{x}$ is the
effect. The equation is deemed causal, because the trajectory, given by%
\begin{equation*}
x(t)=\frac{1}{m}\int_{0}^{t}(t-t^{\prime })F(t^{\prime })\mathrm{d}t^{\prime
}+v_{0}t+x_{0},
\end{equation*}%
clearly demonstrates that $x(t)$ at any time $t>0$ is solely determined by
the force $F(t^{\prime })$ at earlier or equal times $t^{\prime }\leqslant t$%
. To determine the trajectory $x(t)$ in the future up to a time $t_{+}>t$,
we must control, or otherwise know in advance, the force $F(t)$ until that
time. Since this requirement seems to raise no objection, at least in
principle, we conclude that the system is causal.

Instead of $ma=F$, in Dirac's case \cite{DiracRun,calca} we encounter the
equation%
\begin{equation}
m\ddot{x}(t)=\frac{1}{\tau }\int_{t}^{\infty }\mathrm{d}t^{\prime }\hspace{%
0.02in}\mathrm{e}^{\frac{t-t^{\prime }}{\tau }}F(t^{\prime }),  \label{dirac}
\end{equation}%
where $\tau $ is a constant with the dimension of time. This formula is
generated by the parent local, higher-derivative equation%
\begin{equation}
m\ddot{x}(t)-m\tau \dddot{x}(t)=F(t),  \label{force}
\end{equation}%
upon inversion of the operator $1-\tau (\mathrm{d}/\mathrm{d}t)$ through
\textquotedblleft Dirac's prescription\textquotedblright , which requires
perturbativity in $\tau $. The correction appearing in the left-hand side of
(\ref{force}) is the Abraham-Lorentz force, responsible for the infamous
runaway solution, which solves (\ref{force}) but does not solve (\ref{dirac}%
). Thus, Dirac's equation (\ref{dirac}) removes the runaway solution of (\ref%
{force}).

We may assert that equation (\ref{dirac}) violates causality. Indeed, the
solution%
\begin{equation*}
x(t)=\frac{1}{m\tau }\int_{0}^{t}\mathrm{d}t_{1}\int_{0}^{t_{1}}\mathrm{d}%
t_{2}\hspace{0.01in}\int_{t_{2}}^{\infty }\mathrm{d}t_{3}\hspace{0.02in}%
\mathrm{e}^{\frac{t_{2}-t_{3}}{\tau }}F(t_{3})+v_{0}t+x_{0}
\end{equation*}%
shows that, in order to predict $x(t)$, it is not sufficient to know $%
F(t^{\prime })$ at times $t^{\prime }\leqslant t$. Due to the damping factor 
$\mathrm{e}^{(t_{2}-t_{3})/\tau }$, to accurately predict the trajectory up
to time $t$, we require knowledge of $F(t_{3})$ up to times $t_{3}\simeq
t_{2}+\tau $, with $t\geqslant t_{1}\geqslant t_{2}$. Ultimately, this
implies that we need $F(t^{\prime })$ up to times $t^{\prime }\simeq t+\tau $%
.

In other words, if we want to predict the future, we have to anticipate the
external force in a little bit \textit{more} future. Crucially, what happens
within an interval of time of order $\tau $ is \textit{out of our predictive
control}.

Now we highlight the crucial point of our argument: the notion of causality,
as well as its violation, is meaningful only because we label $F$ as an 
\textit{external} force. We cannot apply the same reasoning if the force is
internal, such as one arising from self-interactions.

To illustrate what we mean, consider an elastic force $F=-m\omega ^{2}x$.
Then equation (\ref{dirac})\ becomes%
\begin{equation}
\ddot{x}(t)=-\frac{\omega ^{2}}{\tau }\int_{t}^{\infty }\mathrm{d}t^{\prime }%
\hspace{0.02in}\mathrm{e}^{\frac{t-t^{\prime }}{\tau }}x(t^{\prime }),
\label{didi}
\end{equation}%
and the solution is given by the damped oscillations \cite{calca}%
\begin{equation}
x(t)=c\mathrm{e}^{\lambda t}+c^{\ast }\mathrm{e}^{\lambda ^{\ast }t},
\label{solD}
\end{equation}%
where $c$ is a complex constant, while 
\begin{eqnarray*}
\lambda &=&\frac{1}{3\tau }\left[ 1-\frac{1}{(-1)^{1/3}W}-(-1)^{1/3}W\right]
, \\
W &=&\left( 1+\frac{27}{2}\omega ^{2}\tau ^{2}-\frac{\Upsilon }{2}\right)
^{1/3},\qquad \Upsilon =3\sqrt{3}\,\omega \tau \sqrt{4+27\omega ^{2}\tau ^{2}%
}.
\end{eqnarray*}

Let us pay attention to the following fact: the force on the right-hand side
of (\ref{didi})\ at a time $t$ is determined by the trajectory $x(t^{\prime
})$ at later times $t^{\prime }>t$. This might suggest that causality is
still violated, because \textquotedblleft it is impossible to know the
future in advance\textquotedblright . However, the system is deterministic,
so the future \textit{is}, in fact, known in advance. More accurately, it is
predetermined at all times! Ultimately, the solution depends on two initial
conditions, such as the position $x_{0}=c+c^{\ast }$ and the velocity $%
v_{0}=c\lambda +c^{\ast }\lambda ^{\ast }$ at $t=0$.

Consequently, the model presents no violation of causality when $F$ is
internal to the system. We would not be able to say this if $F$ were
external. The difficulty suggested by the unusual form (\ref{didi}) of the
equation of motion is only apparent, an illusion of acausality that is
unfounded.

We often think that nature is described by the equations of motion (or the
\textquotedblleft physical laws\textquotedblright ) rather than by their
solutions. Instead, the equations belong to our description of nature.
Therefore, it is of little consequence that perfectly valid trajectories
like (\ref{solD}) are generated by equations, such as (\ref{didi}), that
appear acausal to us. In this respect, there is no crucial difference
between equation (\ref{didi}) and a \textquotedblleft
causal\textquotedblright\ equation. In Hume's words, both lead to consistent
\textquotedblleft trains of events\textquotedblright .

This ultimately confirms that the notion of causality loses its meaning in a
deterministic system. It may retain some significance when $F$ is external,
and only as long as such an $F$ is regarded as exempt from the constraints
of determinism. Having noted that quantum theory cannot come to the rescue
here, we need to postulate an $F$ that is not bound to obey the laws of
physics! The question then becomes: what is such an $F$?

We reach similar conclusions in the case of fakeons. The main difference is
that the fakeon equations are supposed to describe fundamental properties of
nature, which means that the challenge they pose to our understanding is
robust, whereas the challenge posed by an effective theory, such as Dirac's
one, can be more easily dismissed as merely apparent. A further key
distinction is that the fakeon equations are symmetric under time reversal,
thus avoiding the need to burden our discourse with an irrelevant arrow of
time.

A typical fakeon equation\ \cite{calca} is%
\begin{equation}
m\ddot{x}(t)=\frac{1}{2\tau }\int_{-\infty }^{\infty }\mathrm{d}t^{\prime }%
\hspace{0.01in}\sin \left( \frac{|t-t^{\prime }|}{\tau }\right) F(t^{\prime
}).  \label{fak}
\end{equation}%
To solve it, knowledge of the external force $F(t^{\prime })$ is required at
all times $t^{\prime }$. However, due to the oscillating behavior of the
sine function, at the practical level it is sufficient to know $F(t^{\prime
})$ only in a neighborhood $|t-t^{\prime }|\lesssim \tau $ of the time $t$
of interest. Since this interval involves some future, we say that (\ref{fak}%
) is acausal.

The parent local, higher-derivative equation is%
\begin{equation}
m\ddot{x}(t)+m\tau ^{2}\ddddot{x}(t)=F(t),  \label{parentf}
\end{equation}%
which gives (\ref{fak}) upon inversion of the operator $1+\tau ^{2}(\mathrm{d%
}^{2}/\mathrm{d}t^{2})$ through the fakeon prescription \cite{calca}.
Equation (\ref{parentf}) has four solutions, of which only two satisfy (\ref%
{fak}).

If we replace the external force with an internal one, such as the elastic
force $F=-m\omega ^{2}x$, the fakeon equation (\ref{fak}) becomes 
\begin{equation}
\ddot{x}=-\frac{\omega ^{2}}{2\tau }\int_{-\infty }^{\infty }\mathrm{d}%
t^{\prime }\hspace{0.01in}\sin \left( \frac{|t-t^{\prime }|}{\tau }\right) 
\hspace{0.01in}x(t^{\prime }),  \label{eqf}
\end{equation}%
which has the solution \cite{calca}%
\begin{equation}
x(t)=c\mathrm{e}^{i\Omega t}+c^{\ast }\mathrm{e}^{-i\Omega t},\qquad \Omega =%
\frac{1}{\tau \sqrt{2}}\sqrt{1-\sqrt{1-4\omega ^{2}\tau ^{2}}},  \label{solf}
\end{equation}%
for $\omega <1/(2\tau )$,~uniquely fixed by the initial position and
velocity. The proper reduction of the set of degrees of freedom to the
physical ones can be proved in fakeon equations with generic
self-interactions \cite{calca}.

Again, we see that the right-hand side of equation (\ref{eqf}) may suggest
acausality, since it implies that the force \textquotedblleft
causing\textquotedblright\ the acceleration on the left-hand side must be
known at all times. However, the force in question is not external, but a
self-interaction. Consequently, it is known at all times \textquotedblleft
for free\textquotedblright\ by solving equation (\ref{eqf})
self-consistently. We thus conclude that no actual violation of causation is
present in this system.

We learn that if we want to introduce a meaningful notion of causality, the
system cannot be independent of its exterior. The idea of cause therefore
does not make sense in an isolated system. It strictly requires the presence
of some truly external entity. Since, however, that entity is necessarily
internal to a larger system, we conclude that causation has no fundamental
meaning in nature.

A popular definition of causality is the requirement that closed timelike
curves (CTCs) are forbidden \cite{Hawking}, on the grounds that such curves
would permit a person to return to their own past by moving forward in time,
thereby gaining the ability to change history. However, as previously
established, if the system is deterministic, changing the past is
impossible, and a CTC provides no help in this respect \cite{novikov}. As
said, one can legitimately claim that the future causes the past, regardless
of whether CTCs exist or not. Moreover, the assertion that \textquotedblleft
one can change one's own past\textquotedblright\ necessarily presupposes a
transcendental power capable of interfering with nature. Then, however, one
need not return to the past to alter it: it is sufficient to change the
future, since this deterministically affects the past. Again, there is no
need of CTCs for this.

Thus, it is evident that CTCs do not violate causality. Unfortunately, the
debate on causality is often burdened by ad hoc definitions and overlaps
with side concepts, leading to claims of causality violations based solely
on breaches of those definitions.

Ultimately, the concept of cause is the result of a sequence of conceptual
missteps and misunderstandings. First, one must erroneously assume that the
world is neither deterministic nor quantum, i.e., that alternative outcomes
are available, but their selection is not governed by quantum chance.
Second, one must postulate something external to nature and endow it with
the \textquotedblleft superpower\textquotedblright\ to make choices
concerning nature. That entity, which would then be designated the
\textquotedblleft cause\textquotedblright ,\ would be deemingly
\textquotedblleft responsible\textquotedblright\ for the effects, in a clear
reflection of the social notions of guilt and responsibility. Third, that
entity acting upon nature would, by definition (because it is supposed to
\textquotedblleft exist\textquotedblright ), have to be part of nature
itself. It would then immediately cease to be a cause, because it would not
be external to nature.

Once again, we observe that the idea of causation is the side effect (!) of
our narrative about nature, but fundamentally does not belong to nature.
Even worse: it belongs to the supernatural, because it postulates the
existence of something external to nature, not subject to the laws of
nature, which can nevertheless act upon nature!

One might counter with the following statement: \textquotedblleft I do not
know whether I will have a cold tomorrow, yet I\ can expose myself to fresh
air, if I\ decide to (or am foolish enough): the cold is the effect, and my
action is the cause\textquotedblright . The weak point of this narration is
the hidden assumption that the \textquotedblleft I\textquotedblright\ is an
entity not subject to the laws of physics, endowed with the mysterious
superpower of acting on nature without being part of it, such as the power
often referred to as \textit{free will}. Once the \textquotedblleft
I\textquotedblright\ is included into a larger system with the rest of
nature (we are made of atoms...), the idea of cause loses its meaning, both
at the classical and quantum levels, for the reasons explained previously.

\section{Prepostdictivity}

\setcounter{equation}{0}\label{prepost}

Causality is commonly believed to grant us the power of controlling and
shaping the future. Having demonstrated that this is a myth, we might still
cultivate the illusion of possessing the less powerful ability of at least
predicting the future. By downgrading the pretense of causality to that of
predictivity, we avoid the necessity of introducing supernatural entities
that are \textquotedblleft external to nature yet capable of acting upon
it,\textquotedblright\ and thus exempt from obeying the laws of physics. The
claim that we can predict nature is less easy to dismiss. However, as we
show in this section, that too is an illusion. For the moment, we adhere to
\textquotedblleft causal\textquotedblright\ equations of motions $ma=F$
(with $F$ external, whatever that might imply). We will relax this
assumption later.

The relevant question for a physicist, then, is: can we make predictions
about the future, relying only upon the present and the past?

\begin{figure}[t]
\begin{center}
\includegraphics[width=16truecm]{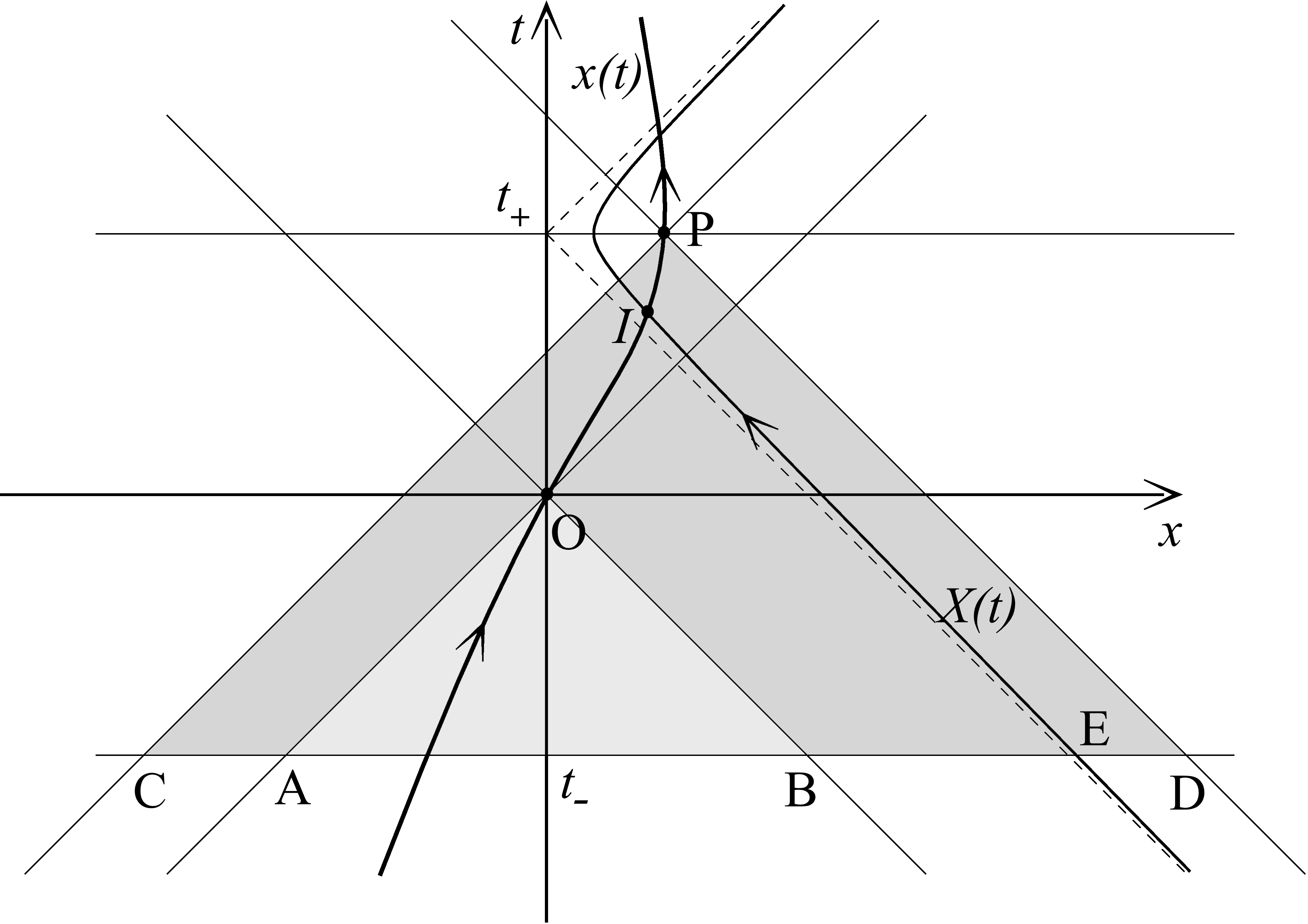}
\end{center}
\par
\vskip -.6truecm
\caption{Prepostidictions}
\label{Contour}
\end{figure}

Our goal is to demonstrate that there exists no physical situation where we
can truly make predictions, that is to say, definitively anticipate the
future within a specified region of space. We first conduct our analysis
within the framework of Special Relativity (SR), and subsequently extend it
to Galilean Relativity (GaR). For simplicity, we restrict our discussion to
one-dimensional systems.

Consider \textquotedblleft us\textquotedblright\ as a point particle with
trajectory $x(t)$, illustrated in fig. \ref{Contour}. At time $t=0$ we are
located at point O and wish to make a prediction concerning the point P at
time $t_{+}>0$.

The maximum knowledge we may possess at O is concentrated in the past
light-cone with tip at O, which we denote by \^{O}. It is important to
stress that we may only have knowledge of objects that we have seen or
detected somehow. We cannot know much about dark\ objects (entities that do
not emit electromagnetic radiation), nor objects that have been obscured by
others. Moreover, we have no knowledge of objects that have never crossed 
\^{O} in their past histories, but cross the past light cone of an event of
our future trajectory, an example being \^{P}, the past light cone with tip
at P.

Consider a body moving along the hyperbolic motion defined by the trajectory 
$X(t)$ shown in the figure: 
\begin{equation*}
x^{2}-(t-t_{+})^{2}=r^{2},\qquad x\geqslant r.
\end{equation*}%
Its past ($t<0$) does not intersect \^{O} (our past light cone at $t=0$).
Its future ($t>0$) crosses \^{P} (the past light cone at the hypothetical
future event P). From O we can have no knowledge of the object in question.
Yet, it can disrupt the future of our system at $I$, the point of
intersection between $x(t)$ and $X(t)$.

We call such an object a \textquotedblleft disruptor\textquotedblright .
When we want to stress that it has no intersection with the past light cone 
\^{O}, we call it an \textquotedblleft extradisruptor\textquotedblright .

We conclude that, even assuming that we live in a deterministic universe, at
time $t=0$ we cannot be certain that we will end at the point P at time $%
t_{+}$, as predicted by the equations of motion in the absence of
disruptors. The possibility of extradisruptors, objects unknowable from our
past light cone \^{O}, but capable of influencing our future trajectory,
fundamentally implies that we cannot make definitive predictions about the
future.

Assume that we started collecting data at some time $t_{-}<0$. Our maximum
knowledge at $t=0$ relies on what intersected \^{O}. Extradisruptors are, by
definition, out of reach, as a matter of principle. In practice we cannot
even exclude \textquotedblleft intradisruptors\textquotedblright , such as
dark or shadowed objects that intersected \^{O} but did not send any or
enough signals towards us.

Therefore, how can we fix the initial conditions at, say, time $t_{-}$? The
past light cone \^{O} is a proper subset of \^{P}. Consequently, its
intersection with the horizontal constant-time slice $t=t_{-}$ spans a
smaller segment (AB) compared to the intersection of \^{P} with the same
slice, which spans a larger segment (CD). We may have knowledge of objects
that intersected \^{O}, sent signals to us, then exited \^{O} and
subsequently reached the segment AB or CD. However, because we lack any
knowledge of disruptors, we have no way to fix complete initial conditions
in CA\ and BD\ at times $t\leqslant 0$. The conclusion is that we cannot
predict what is going to happen to \textquotedblleft us\textquotedblright ,
or our system, at a later time $t_{+}$.

For the sake of completeness, let us consider the limit $t_{-}\rightarrow
-\infty $. We have the following paradox: every $x$ appears to be in \^{O},
because for every $x$ there is a remote time before which $x$ belongs to 
\^{O}. Nevertheless, extradisruptors have no intersection with \^{O}, which
means that points $x$ outside \^{O} (i.e., those belonging to the complement
of \^{O} inside \^{P}) must exist. What is important here is that not even
taking $t_{-}=-\infty $ allows us to gather enough initial conditions to
predict the future beyond O. Even in the best-case scenario, extradisruptors
cannot be excluded.

Let us examine potential ways to circumvent this difficulty. An option is to
assume that the system is isolated. As common as this assumption is, it is
not realistic, as we cannot build walls or shields sufficiently robust to
ensure that no external object, like the one with trajectory $X(t)$, can
disrupt the experiment. Another possibility is to assume that nothing else
exists in the universe besides the point particle with trajectory $x(t)$.
This, however, is an unjustified idealization, and a metaphysical stretch,
since it amounts to making assumptions about regions of spacetime that are
fundamentally inaccessible to us without invoking superluminality.

Ultimately, the best we can do is make our \textquotedblleft
prediction\textquotedblright , hope that no disruptions, like the ones
previously described, occur in the meantime, and check a posteriori that the
outcome in P is the predicted one. Then, and only then, can we assert that 
\textit{we were able} to predict it. However, this retrospective
verification process is fundamentally a postdiction, which is why our
initial statement\ was not a prediction, but rather a \textquotedblleft
prepostdiction\textquotedblright .

At first, one might assume that the limitations just described are inherent
to Special Relativity, hoping that they are absent in nonrelativistic
mechanics. This is incorrect, as analogous difficulties persist in the limit 
$c\rightarrow \infty $, where SR reduces to Galilean Relativity. In this
limit, the past light cones depicted in the figure become half-planes, yet
the knowledge available at O is still insufficient to determine the future
at P. In GaR, macroscopic bodies can move at arbitrarily high speeds, making
it impossible to exclude that a heavy item could traverse a very large
distance during the time interval between $t=0$ and $t_{+}$, thereby
disrupting the outcome at P, regardless of the robustness of any wall built
to supposedly isolate the system. To fully exclude such a possibility, we
would need to specify the initial conditions of every object in the
universe, which is patently impossible. Once again, we are left to hope for
the best, bet that no disruption occurs in the meantime, and check a
posteriori at P whether we got away with it or not. Thus, even in the
nonrelativistic limit, we can only make prepostdictions.

\subsection{Disruptors, incoming waves and statistics}

An objection might be raised: why has no unfortunate situation like those
feared here ever disrupted any experiment\footnote{%
A physical scenario that may illustrate the failure of prediction is the
phenomenon of rogue waves (which may be seen as intradisruptors) and
possibly tsunamis (extradisruptors) in a stormy sea.}? If disruptions are so
statistically disfavoured, why should we bother? Isn't discarding such
occurrences equivalent to assuming standard boundary conditions at infinity
in classical electrodynamics (i.e., postulating that there are no sources at
infinity sending signals towards us)? Otherwise we would not be able to
trust solutions that rely solely on the retarded potentials. Isn't it
paranoid to suspect that nature is conspiring from far away to mess with us
by disrupting our experiments \textquotedblleft on purpose\textquotedblright
?

The core point is that we are questioning a candidate \textit{physical} 
\textit{principle} here: the ability to truly predict in physics. If a
statement holds true only barring extremely rare and unlikely situations, it
cannot be elevated to the rank of a fundamental principle. While conceptual
shortcuts are more than sufficient for everyday life and many common
physical situations, they are not guaranteed to be adequate when inquiring
about quantum gravity, the microscopic world, or, more generally, the
unknown. This distinction constitutes the rationale of our investigation. We
are questioning whether the methodology usually adopted is strategically
sound for research on fundamental interactions. Our suggestion is that it is
not.

Returning to the common, and apparently innocuous, assumption of
\textquotedblleft no incoming radiation from infinity\textquotedblright ,
which dictates the use of the retarded potentials in classical
electrodynamics, this choice concerns regions of spacetime that are
inaccessible to us, even in theory. Consequently, the adoption of the
retarded solution, such as (\ref{Am}), over the advanced one, such as (\ref%
{Ap}), relies on an unprovable assumption: in effect, a \textquotedblleft
no-disruptors\textquotedblright\ bet, and a metaphysical stretch.

Every light signal perceived by our eyes is a potential disruptor. Indeed,
1) it originates from regions of spacetime that were inaccessible to us
prior to the moment of perception; and 2) its source may have remained
outside our past light cone until that very moment. Is it possible to gather
enough knowledge to predict with absolute certainty, at least in theory,
what we will see in a minute from now? The answer is no, and not merely for
practical reasons, but as a matter of principle. All we can do is wait and
see.

Under \textquotedblleft normal\textquotedblright\ circumstances, the
retarded choice is natural at the macroscopic level. It is statistically
implausible that, once an oscillating dipole\ such as (\ref{source}) is
turned on, a coherent influx of waves coming from infinity would conspire to
transform it into a \textquotedblleft sink\textquotedblright . This is the
key point of our argument: the choice between (\ref{Am}) and (\ref{Ap}) is
dictated by statistics; it is not a matter of cause and effect.

Let us dig more into this. At our scales, the retarded pair (\ref{source})-(%
\ref{Am}) describes emission, whereas the advanced pair (\ref{source})-(\ref%
{Ap}) describes reception\footnote{%
Actually, reception is described by an incident wave that strikes the
smartphone and propagates beyond it, rather than an incoming wave. This
distinction is not crucial for our discussion.}. Upon our input, a
smartphone generates a motion of charged particles, as in (\ref{source}), to
encode, for instance, a speaking voice. This produces a field like (\ref{Am}%
), that is, the emission of a signal for communication. Conversely, an
incoming field such as (\ref{Ap}) reaches the smartphone from an external
source, independent of any input from us, and induces a motion like (\ref%
{source}) of charged particles within the antenna. The device then
translates this signal into data, a voice, or a message. The input is a
potential disruptor, since it originates from a region of spacetime that
remained inaccessible until the very moment the phone began to ring.

Now, consider a \textquotedblleft mirror\textquotedblright\ universe
characterized by inverted boundary conditions at infinity and a statistical
framework entirely reciprocal to our own, where highly improbable events
manifest consistently, coincidence after coincidence, over vast timescales%
\footnote{%
The mirror\ scenario we are describing could just be an extraordinary
statistical fluctuation within our own universe.}. For instance, an incoming
field such as (\ref{Ap}) would be coupled with a deliberate action, such as
the oscillation (\ref{source}) of charged particles triggered by our
command; notably, this process would result in a net gain of energy rather
than its expenditure. Conversely, whenever an outgoing field like (\ref{Am})
occurs, it corresponds to a sudden, unplanned oscillation like (\ref{source}%
), triggered by thermal noise.

Should we conclude that (\ref{Ap}) is the cause of (\ref{source})? If so, it
would imply that our will is determined by the incoming field. In this view, 
\textit{free} will would not exist: human beings would be governed by nature
like automata\footnote{%
Note that we would have no possibility to \textquotedblleft change our
mind\textquotedblright\ after gaining knowledge about an incoming wave.
Specifically, assume that Bob is positioned at $r>0$ and perceives an
incoming wave at time $t-r$ converging towards the origin. Bob then knows
that someone there, say Alice, will activate $J^{\mu }$ at time $t$ (given
our assumption of inverted statistical laws). However, Bob cannot reach
Alice in time to intervene, as the crucial information would need to travel
faster than light to get to\ her before she enacts her intention. In other
words, the incoming wave acts as an extradisruptor for Alice. Alice and Bob
can only meet later and share their story, at which point Alice would
discover that she has no free will, since Bob knew her plans in advance.}.
Alternatively, one could insist that \textquotedblleft (\ref{source}) is the
cause of (\ref{Ap})\textquotedblright , that is, not only do we possess free
will, but it is so potent that it deterministically alters the past. From a
purely physical standpoint, the choice of narrative is indifferent.

It must be emphasized that this argument hinges on the presence of a
\textquotedblleft subject\textquotedblright . Otherwise we would be unable
to distinguish \textquotedblleft planned\textquotedblright\ or
\textquotedblleft wanted\textquotedblright\ actions from \textquotedblleft
unplanned\textquotedblright\ ones, rendering the discourse even more
tenuous. Again, we see that notions such as cause, or free will, cannot be sustained
unless one invokes either metaphysics (the subject), or statistics (the law
of large numbers, or the boundary conditions at infinity).

The important point is that the specification of boundary conditions at infinity
involves regions of spacetime that are inaccessible without invoking superluminality. 
The choice we usually opt for at ordinary scales is driven by statistical properties inherent to the
macroscopic world, rather than by a fundamental property of nature.  
Hence, extending that choice, or an equivalent one, to the microscopic world is unjustified.
There is no reason why the universe at infinitesimal scales should resemble
the universe at large scales (which means: a desert expanse, where the
assumptions of no incoming radiation and no disruptors feel natural),
instead of, say, a turbulent, stormy sea. For example, fakeons, through
equations like (\ref{fak}), contemplate a superposition of incoming and
outgoing waves at the microscopic level, precisely recalling a stormy sea.

The conclusion is that the arrow of time resulting from causality (possibly
downgraded to predictivity)\ resembles the thermodynamic arrow of time, in
the sense that both rely on the law of large numbers. In the same way as
statistical irreversibility does not hold at the microscopic level (given
that the law of large numbers breaks down with small numbers), causation and
predictivity lose meaning at small distances (even in the presence of
external forces), because assumptions barring disruptors or incoming waves,
which rely on statistics, cannot be trusted there. These arguments point
towards causation and predictivity as emerging properties of (our
description of) the universe rather than fundamental ones, thereby making
the abandonment of microcausality and micropredictivity not only acceptable,
but also necessary.

In passing, it is intriguing to note that fakeon theories imply that the
universe is endowed with a \textquotedblleft radial arrow\textquotedblright\
pointing from the microscopic \textquotedblleft stormy
sea\textquotedblright\ to the macroscopic \textquotedblleft desert
expanse\textquotedblright . If fakeon models describe reality, this is a
rigorous arrow, not just an approximate one.

\subsection{Remarks}

We end this section by addressing other, minor issues on prepostdictivity.
Disruptors are part of the \textquotedblleft external
forces\textquotedblright\ $F$\ that enter the equations of motion $ma=F$.
Hence, it may be objected that we can still describe the disruptions,
whenever they occur, by means of ordinary physical laws. The key point is
that we can only describe them a posteriori, but we cannot anticipate them a
priori. External forces like these cannot be arranged as we wish, controlled
or predicted. Was not the ability to control nature the rationale for
postulating causes and \textquotedblleft external\textquotedblright\ forces
in the first place?

A possibility is to downgrade the predictions to statistical ones, in the
following sense. If the same experiment is repeated several times, the
disruptions will not be the same, which allows us to filter their effects
away, so to speak. Yet, how do we correctly average over them? What exactly
constitutes an \textquotedblleft average prepostdiction\textquotedblright ?
How can we quantify the statistics of disruptions without making assumptions
about unreachable regions of spacetime? Besides, when we resort to
statistical predictions, as we do in quantum mechanics, it is precisely
because we cannot predict the outcome of a single experiment.

Summarizing, when we downgrade causation to mere chronological ordering, we
are essentially demoting the major purpose of controlling, or shaping, the
future evolution of a system to the more modest goal of predicting it.
However, as demonstrated, it is unjustified to elevate predictivity to a
fundamental principle, or extend it to the microscopic world. A fortiori,
the same conclusion applies to causality.

The impossibility of predicting the future cannot be entirely mirrored onto
the impossibility of tracing back the past. It is true that if we begin
collecting data at some time $t=t_{+}$ at point P, we cannot know whether
intradisruptors have changed the course of events at $t<t_{+}$. However, we
cannot mirror extradisruptors, since they always intersect past light cones
lying in the future. This means that we may have knowledge of them when we
trace back the past, at least in principle. To some extent, this introduces
a sort of arrow pointing from the present to the future. However, this is
still not a fundamental property of the universe. It merely reflects our
ability to keep memory of events, a possibility that arises when macroscopic
or non isolated systems are involved. At the microscopic level we have
perfect symmetry, as shown in the example of Section \ref{det} concerning
spin measurements along the $x$ and $z$ directions. Besides, if we want a
system to be visible to us, it cannot be isolated, since it must radiate
towards us.

\section{Delayed prepostdictivity}

\setcounter{equation}{0}\label{delprepost}

In\ the previous section we argued in the realm of \textquotedblleft
causal\textquotedblright\ equations of motion $ma=F$. What about
prepostdictivity in fakeon equations like (\ref{fak}), or Dirac's equation, (%
\ref{dirac})?

In \textquotedblleft plain\textquotedblright\ prepostdictivity (the one
associated with $ma=F$), we need to know the external force $F$ up to the
moment $t_{+}$ of our \textquotedblleft prediction\textquotedblright . We
have established that, since $F$ is beyond our control for the reasons
explained before, we cannot really predict. We can at most cross our fingers
and prepostdict. Yet, we do not need to know $F$ \textit{beyond} the time $%
t_{+}$.

On the other hand, in systems governed by nonlocal equations like (\ref{fak}%
) and (\ref{dirac}), we need to know $F$ in a little bit more future, which
is approximately $t_{+}+\tau $, where $\tau $ is the characteristic scale of
nonlocality. Thus, at time $t_{+}$ we cannot postdict what happened till
then. We can at most postdict what happened till $t_{+}-\tau $. To postdict
till $t_{+}$, we have to be more patient and wait till $t_{+}+\tau $. This
means that if we are making a statement at time $t=0$ that aims to reach as
far as $t_{+}$, it is not enough to cross our fingers till $t_{+}$. We need
to cross our fingers till $t_{+}+\tau $.

The outcome is that in fakeon systems, as well as in Dirac's one, we do not
have plain prepostdictivity, but rather a \textit{delayed} prepostdictivity,
with the delay being quantitatively defined by the characteristic time scale 
$\tau $.

We can illustrate this fact more explicitly under the assumption that $\tau $
is small. Then equations (\ref{dirac})\ and (\ref{fak})\ can be written as%
\begin{eqnarray*}
\text{Dirac}\text{:\qquad } &&m\ddot{x}(t)=F(t+\tau )+\mathcal{O}(\tau
^{2})=F(t)+\Delta _{\tau }F(t)+\mathcal{O}(\tau ^{2}), \\
\text{fakeon}\text{:\qquad } &&m\ddot{x}(t)=F(t)-\Delta _{\tau }^{2}F(t)+%
\mathcal{O}(\tau ^{4}),
\end{eqnarray*}%
respectively, where $\Delta _{\tau }F(t)=F(t+\tau )-F(t)$ is the forward
difference, while $\Delta _{\tau }^{2}F(t)=F(t+\tau )-2F(t)+F(t-\tau )$
denotes the second central difference. We see that the right-hand sides
explicitly involve the force in the future time $t+\tau $.

In passing, we note that a truncated equation such as $\ddot{x}(t)=-\omega
^{2}x(t+\tau )$ (a harmonic oscillator with a delayed elastic force) admits
infinitely many solutions, as is generally expected of nonlocal
formulations. Instead, the Dirac and fakeon equations, while nonlocal, are
of special types, being tied to parent local equations. Their solution space
is just the physically expected one \cite{calca}.

Coming back to our problem, a possible way out is to interpret $\tau $ as
the minimum time resolution, that is to say, to postulate that we cannot
experimentally distinguish events that are separated by time intervals
shorter than $\tau $. The \textquotedblleft present\textquotedblright\ is
therefore defined as \textquotedblleft present within an interval of time of
order $\tau $\textquotedblright . In this scenario, $t$ and $t+\tau $ can be
regarded as \textquotedblleft the same time\textquotedblright , and the
equations may be considered causal.

Even when we postulate a truly external force $F$, the violation of
causality fades away as soon as we interpret $\tau $ as the minimum time
resolution. This means that, strictly speaking, we cannot definitely claim
that the Dirac and fakeon equations imply the violation of causality, not
even in the presence of external entities: they may merely imply the
existence of a fundamental limit on temporal resolution.

The crucial difference is that Dirac's equation is effective, so the minimum
time resolution $\tau $ it contains cannot hint to a fundamental property of
nature. In the fakeon case, $\tau $ may suggest a fundamental impossibility
of distinguishing events separated by time intervals shorter than $\tau $.
However, as we now demonstrate, neither is this true.

It should be noted that the interpretation of $\tau $ as the minimum time
resolution makes sense only if the force $F$ is truly external. If $F$ is a
self interaction, there is no problem in interpreting solutions such as (\ref%
{solD}) and (\ref{solf}) for arbitrarily short intervals of time. Given that
any external force is internal to a larger system, the fakeon models do not
predict a \textit{fundamental} time resolution $\tau $ in the universe.
Neither does Dirac's system, because it is inherently an effective model.

Thus, even if we adhere to the premise that there exist entities external to
nature that can act on nature, it is still not true that fakeon theories
violate causality, unless we assume that we can experimentally measure (and
thereby give physical sense to) arbitrarily short intervals of time. Indeed,
as soon as we accept the possibility that time in nature is equipped with a
minimum time resolution $\tau $ (applicable only to those external
entities), discussing violations of causality below $\tau $ becomes
meaningless.

Ultimately, we reach a situation where the limitations represented by $\tau $
concern the external force, and not nature itself. Even if we hypothesize
the existence of entities external to nature but endowed with the superpower
of acting upon it, quantum gravity with fakeons compels us to accept
limitations on their power (whether it is a delay of prepostdictivity or a
minimum time resolution). In other words, nature, i.e., an \textquotedblleft
infrapower\textquotedblright , is able to place limits on a supposed
\textquotedblleft superpower\textquotedblright\ transcending nature. A
physical limitation on something that does not even exist physically, but
merely encodes our pretense of controlling nature, is not a huge price to
pay, especially if the reward is a testable theory of quantum gravity.

\section{Causality in quantum field theory}

\setcounter{equation}{0}\label{cauqtf}

The two main definitions of causality in quantum field theory are
Bogoliubov's condition and the Lehmann-Simanzik-Zimmermann (LSZ) formulation 
\cite{diagrammar}.

Bogoliubov's condition \cite{bogoliubov} applies to diagrams and off-shell
correlation functions. In simplified terms, it requires that if a spacetime
point $x_{1}$ is in the timelike future of $x_{2}$, diagrams involving both $%
x_{1}$ and $x_{2}$ can be arranged so that energy flows only forward in
time, from $x_{2}$ to $x_{1}$.

The problem with this definition is that one cannot accurately assign
spacetime points to on-shell particles, because relativistic wave packets
spread, thus obscuring the notion of a definite particle position. Hence,
Bogoliubov's causality cannot be formulated as a constraint on the S matrix
itself, but merely applies to off-shell correlation functions. These are, in
general, gauge dependent. Although they can be made gauge invariant (even
for insertions of elementary fields, by working with physical degrees of
freedom only), this may come with the price of introducing nonlocalities
(see next section), which obscure a direct spacetime cause $\rightarrow $
effect picture.

The LSZ definition \cite{LSZ} of\ causality is enforced via field
commutators, which are required to vanish for spacelike separation. When
fields are expressed in terms of diagrams, this condition reduces precisely
to Bogoliubov's energy-flow condition. From our perspective, working with
fields is no different than working with off-shell correlation functions.
Consequently, the conceptual difficulties associated with the gauge
dependence and the obscurity of a direct spacetime cause $\rightarrow $
effect picture remain fully relevant even under the LSZ formalism.

The main weakness of these conditions, and\ of alternatives that have been
proposed in the literature, is that they sound merely technical, or even
artificial. It is hard to relate them to an intuitive notion of causation,
which is itself problematic for the reasons explained in the previous
sections. Aside from the risk of overlapping causality with non
superluminality, the proposed notions often become entangled with locality
and analitycity, whereas one needs to keep such properties distinct from one
another, especially because analyticity is not a physical principle and
locality is mostly a convenient requirement. There is no doubt that the
Bogoliubov and LSZ conditions imply constraints on physical quantities, but
they offer no clear motivation why nature should conform to those.
Ultimately, causality in QFT sounds like a condition on the mathematical
scaffolding, not on the physical predictions.

In quantum gravity with fakeons, the Bogoliubov and LSZ conditions are
expected to be violated at the microscopic level. These violations are the
quantum counterparts of the requirement that the external force $F(t)$ be
known in a little bit more future, according to the classicized equation (%
\ref{fak}), as discussed in the past section. The outcome, which we
reiterate here, is that if fakeons participate in the description of nature
in any way, they put limits on the illusory superpower of controlling, or
even just predicting, nature. The apparent paradox that an infrapower can
constrain a superpower evaporates as soon as we accept the obvious fact that
nothing can act on nature without being part of it, thereby collapsing the
Platonic illusion of a world of ideas existing \textquotedblleft
above\textquotedblright\ or even just \textquotedblleft
outside\textquotedblright\ nature, yet capable of acting on it (as
\textquotedblleft souls\textquotedblright , for example).

One could say that quantum gravity cuts to the chase and tolls the death
knell for causation, but one could also argue that the QFT framework of the
Standard Model of particle physics had already planted the seed of doubt so
deeply that quantum gravity does not actually add much. Both positions sound
well motivated, with the caveat that fakeons may have a stronger impact in
prompting one to pause and delve into the issue profoundly enough to settle
the matter once and for all.

\section{Nonlocality in gauge theories, gravity and fakeon models}

\setcounter{equation}{0}\label{loc}

The other unusual feature of equations like (\ref{dirac})\ and (\ref{fak})\
is their nonlocality, which may motivate some criticism by those who believe
in some form of locality as a fundamental principle.

It should be noted that, more than a principle, the requirement that the
classical Lagrangian should be local in quantum field theory is merely a
recipe that has worked successfully so far, especially in conjunction with
renormalizability. Minor tweaks to this assumption are not anticipated to
pose dramatic risks.

We distinguish between hard nonlocality and soft nonlocality. Hard
nonlocality refers to theories, such as those pursued by Krasnikov, Kuz'min,
Tomboulis, Modesto and others \cite{nonlocal}, where the classical
Lagrangian is genuinely nonlocal. Soft nonlocality refers to the nonlocality
encountered in fakeon theories. Crucially, in both cases, the nonlocality is
sufficiently restricted to ensure that the equations of motion are not
burdened with the need to specify infinitely many initial conditions: the
standard physical initial conditions suffice \cite{calca}.

The fakeon models do not possess a standard classical Lagrangian. One can
derive a \textquotedblleft classicized\textquotedblright\ Lagrangian (which
is nonlocal) from a parent, local Lagrangian by integrating out the fakeon
fields with the appropriate prescription. An example of classicized
Lagrangian is (\ref{milan}) below, derived from (\ref{inter}).

The descent is a projection that recalls (with due differences) the
projection of the Lagrangian in gauge and gravity theories, obtained upon
elimination of the unphysical degrees of freedom (which are: the temporal
and longitudinal modes of the gauge fields; those of the fluctuation of the
metric around flat space; the trace of the fluctuation of the spatial
metric; and the Faddeev-Popov ghosts). The result is nonlocal in those
familiar cases as well, even in the classical limit, although this fact
mostly goes unnoticed.

Consider, for example, classical electrodynamics. The Lagrangian%
\begin{equation*}
\mathcal{L}_{\text{QED}}=-\frac{1}{4}F^{\mu \nu }F_{\mu \nu }+\bar{\psi}%
[i\gamma ^{\mu }(\partial _{\mu }+ieA_{\mu })-m)]\psi
\end{equation*}%
is local, but contains fields that do not correspond to physical degrees of
freedom. If we choose a simple gauge such as $A_{L}=0$, for some
longitudinal component $A_{L}$, we obtain%
\begin{eqnarray*}
\mathcal{L}_{\text{QED}}^{\prime } &=&\frac{1}{2}(\partial _{\mu }\mathbf{A}%
_{\perp })(\partial ^{\mu }\mathbf{A}_{\perp })+\frac{1}{2}(\mathbf{\partial 
}_{\perp }\cdot \mathbf{A}_{\perp })^{2}+\bar{\psi}(i\gamma ^{\mu }\partial
_{\mu }-m)\psi +\mathbf{J}_{\perp }\cdot \mathbf{A}_{\perp } \\
&&-\frac{1}{2}A_{0}\Delta A_{0}-(\mathbf{\partial }_{\perp }\cdot \mathbf{A}%
_{\perp })(\partial _{0}A_{0})-\rho A_{0},
\end{eqnarray*}%
which is still local (at least in a suitable frame, such as $A_{L}=A_{3}$),
but explicitly contains the unphysical (nonpropagating) component $A_{0}$ of
the vector potential, besides the transverse components $\mathbf{A}_{\perp }$%
. Here $\rho =e\psi ^{\dagger }\psi $ is the charge density and $\mathbf{J}%
_{\perp }=e\bar{\psi}\mathbf{\gamma }_{\perp }\psi $ is the transverse
current.

Integrating out $A_{0}$ gives the truly nonlocal, classical Lagrangian 
\begin{eqnarray*}
\mathcal{L}_{\text{QED}}^{\prime \prime } &=&\frac{1}{2}(\partial _{\mu }%
\mathbf{A}_{\perp })(\partial ^{\mu }\mathbf{A}_{\perp })+\frac{1}{2}(%
\mathbf{\partial }_{\perp }\cdot \mathbf{A}_{\perp })^{2}+\bar{\psi}(i\gamma
^{\mu }\partial _{\mu }-m)\psi +\mathbf{J}_{\perp }\cdot \mathbf{A}_{\perp }
\\
&&+\frac{1}{2}(\mathbf{\partial }_{\perp }\cdot \mathbf{\dot{A}}_{\perp
}-\rho )\frac{1}{\Delta }(\mathbf{\partial }_{\perp }\cdot \mathbf{\dot{A}}%
_{\perp }-\rho ),
\end{eqnarray*}%
which contains the physical degrees of freedom only and demonstrates the
necessity of nonlocal terms at the classical level when unphysical modes are
eliminated.

Similarly, in theories with fakeons the classicized, nonlocal Lagrangian is
the projection of a parent local one by means of the fakeon prescription.
For example, consider a simple model with the parent local Lagrangian \cite%
{calca} 
\begin{equation}
\mathcal{L}=\frac{1}{2}(\partial _{\mu }\varphi )^{2}-\frac{m^{2}}{2}\varphi
^{2}-\frac{1}{2M^{2}}\phi \left[ (\Box +m^{2})^{2}+M^{4}\right] \phi -\frac{g%
}{2}\phi \varphi ^{2},  \label{inter}
\end{equation}%
where $\varphi $ is a physical field and $\phi $ is an extra field that we
want to quantize as a fakeon. Integrating out $\phi $ in the appropriate
way, we obtain the classicized, nonlocal Lagrangian 
\begin{equation}
\mathcal{L}_{\text{cl}}=\frac{1}{2}(\partial _{\mu }\varphi )^{2}-\frac{m^{2}%
}{2}\varphi ^{2}+\frac{g^{2}}{8}\varphi ^{2}\left. \frac{M^{2}}{(\Box
+m^{2})^{2}+M^{4}}\right\vert _{\text{f}}\varphi ^{2}\,,  \label{milan}
\end{equation}%
where the subscript \textquotedblleft $\text{f\textquotedblright }$ stands
for the fakeon Green function (see \cite{calca}).

This part of the discourse does not extend \textit{verbatim} to Dirac's
case, where a local Lagrangian for (\ref{force}) is not available. Yet, the
manipulations described earlier at the level of equations of motion (\ref%
{dirac}) and (\ref{force}) are sufficient to convey the intended idea.

The other point is that, although there exist local, gauge invariant
observables in gauge and gravity theories, they do not form a basis. A basis
must include nonlocal observables. For example, in QED\ we have local
observables such as $F_{\mu \nu }$, $\bar{\psi}\psi $, $F^{\mu \nu }F_{\mu
\nu }$, etc., but no local \textquotedblleft electron
observable\textquotedblright . Yet, we know that we can observe the
electron. To explain this, we need to switch to nonlocal observables, such as%
\begin{equation}
\exp \left( -ie\frac{1}{\Delta }(\mathbf{\nabla }\cdot \mathbf{A})\right)
\psi ,  \label{did}
\end{equation}%
a trick again suggested by Dirac \cite{diraccloud}. More generally (as
discussed in refs. \cite{offshell}), we can consider expressions like%
\begin{equation}
\exp \left( ie\left. \frac{1}{\Box }\right\vert _{\text{f}}(\partial ^{\mu
}A_{\mu })\right) \psi ,  \label{dif}
\end{equation}%
where the fakeon prescription is used to invert the D'Alembertian. The
reason for preferring (\ref{dif}) over (\ref{did}) is that (\ref{did})
explicitly breaks Lorentz invariance. At the same time, simply replacing the
Laplacian $\Delta $ with minus the D'Alembertian $\Box $ (and $\mathbf{%
\nabla }\cdot \mathbf{A}$ with $\partial ^{\mu }A_{\mu }$) would introduce
unphysical degrees of freedom if a different prescription were used for the $%
\Box ^{-1}$ operator. The fakeon approach to gauge-invariant observables can
be generalized to non-Abelian theories and gravity (see \cite{offshell}
again).

Thus, we can assert that gauge and gravity theories are inherently nonlocal.
This realization makes it harder to dismiss the extra nonlocality brought in
by fakeons. The two types of nonlocality are actually in the same class,
since both descend from a parent locality through projection.

We recall that, while unitarity in gauge and gravity theories is ensured by
the Ward--Takahashi--Slavnov--Taylor identities following from the local
symmetries, unitarity in fakeon models follows from a modified diagrammatics 
\cite{diagrammarMio}, based on non-time-ordered correlations functions \cite%
{PVP20}. The lack of time ordering at high energies is another (expected)
facet of the issues discussed here.

\section{Conclusions}

\setcounter{equation}{0}\label{concl}

The debate on causality in physics is a typical example of a false start,
i.e., taking the concept for granted without adequately investigating it
first, and spinning the discussion around it endlessly. In this paper we
have considerably extended the widespread skepticism surrounding the notion
of cause, driven by the developments brought about by quantum gravity.

The key question is: what constitutes a \textquotedblleft
cause\textquotedblright ? We have argued that causation can only have a
meaning when it refers to entities that are external to the physical system
under observation. By this we mean external in an extreme sense:
\textquotedblleft external to physics\textquotedblright , since they are not
allowed to obey physical laws, otherwise they would be internal to a larger
system and would consequently lose their essence as \textquotedblleft
causes\textquotedblright . Examples are concepts like free will or soul.
Pushing this argument further, it becomes clear that the idea of cause
belongs to metaphysics, or the transcendental, precisely like the notions of
free will and soul, hence it should be abandoned in fundamental science.

First quantum mechanics, then quantum field theory, and, more recently,
quantum gravity, have fueled the need to place the notion of cause under
close scrutiny. Quantum field theory poses nontrivial challenges to defining
causation in a satisfactory way, and quantum gravity, specifically the
fakeon framework, cuts to the chase by dismantling the notions of
\textquotedblleft event\textquotedblright\ and chronological ordering for
sufficiently short intervals of time, even in the presence of hypothetical
truly external forces.

If controlling nature and making it do what we want is too much to ask, let
us content ourselves with predicting it. Or should we? Hume argued that
(conveying his central argument, though not in his exact words) the mere
observation that a freely falling body has always fallen to the ground up to
the present day does not provide logical certainty that it will do so
tomorrow. In other words, the laws of physics are not truly
\textquotedblleft laws,\textquotedblright\ but rather postdictions (if
viewed from the future to the past) and bets (if viewed from the past to the
future). We bet that tomorrow a free body will still fall to the ground. If
it will not, we will (opportunistically) modify the physical
\textquotedblleft law\textquotedblright\ by adding a correction to account
for the new effect, whatever it may be. From that moment onward, we will
make refined bets by incorporating the correction. There is nothing in this
way of proceeding --- which is, in fact, how we proceed --- that allows us
to elevate our findings to \textquotedblleft principles\textquotedblright\
or physical \textquotedblleft laws\textquotedblright\ and be confident that
nature will conform to \textquotedblleft our\textquotedblright\ laws
tomorrow in the same way as it did so far.

Even in this context, we extended Hume's skepticism significantly. Suppose
that the physical laws as we know them are indeed laws that bind nature
tomorrow and forever. We have argued that even then we cannot make
predictions about the future. Not only, but we have shown that this follows
from the supposed \textquotedblleft laws\textquotedblright\ themselves! We
can never exclude that disruptors\ will appear \textquotedblleft out of
nowhere\textquotedblright , so to speak, and change the final outcome.
Systems cannot be sufficiently isolated, nor can initial conditions be
completely fixed. These limitations hold even without questioning the future
validity of the so-called physical laws.

Consequently, physicists can only make prepostdictions (requiring
retrospective verification) at large scales, relying on the law of large
numbers to dismiss potential disruptions. This confirms that the illusory
arrow of time associated with causality is inherently statistical, and
emerges solely at the macroscopic level, thereby rendering microcausality
unwarranted and strained.

Not much of this changes when quantum gravity is included, where we can at
most make \textit{delayed} prepostdictions. In theories with fakeons, the
delay is the reciprocal of the fakeon mass $m_{\chi }$ (about 10$^{-37}$
seconds), short enough to be consistent with observation \cite{Marino}.
Ultimately, the further renunciation entailed by quantum gravity with
fakeons is not overly demanding.

Since the birth of science, physicists have lived under the mirage that they
could predict \textit{something}. This is an illusion that has never truly
held. The truth is that we just place bets and cross our fingers: a
posteriori, we can verify that events unfolded as expected, but we have no
way of guaranteeing \textit{a priori} that it will indeed be the case.

To some, these might sound like matters of hairsplitting, but confronting
the problem of quantum gravity is precisely the kind of challenge where
splitting hairs may truly matter. Others may think that the positions
expressed here are \textquotedblleft extreme\textquotedblright , but we have
emphasized that what is truly extreme is postulating the existence of
transcendental entities (such as the so-called \textquotedblleft
causes\textquotedblright ) that are outside nature, yet endowed with the
superpower of acting on nature (setting aside the absurdity of assuming the
\textquotedblleft existence\textquotedblright\ of those entities, given that
what \textquotedblleft exists\textquotedblright\ is part of nature by
definition). Brushing these issues aside and continuing research as before
comes with the risk of overlooking profound opportunities. It is better to
accept that quantum gravity may require walking on thin ice, and cope with
the fact that exploring the unknown may well demand that we challenge even
the most basic principles\ we have long taken for granted.

String theory, for example, is claimed to be strictly causal. One of its
many flaws is that in order to reproduce all of its vibrational modes, it
effectively contains an infinite tower of particle states. Another is that
its standard formulation is inherently restricted to computing on-shell
scattering amplitudes between asymptotic states. After decades, no accepted
formalism to describe off-shell quantities is available. A third flaw is its
lack of predictivity. These are huge prices to pay to perpetuate a
controversial \textquotedblleft principle\textquotedblright\ (perhaps even a
stereotype) such as causality. And why should we be willing to pay that
price? To maintain the illusion that we are external to nature and possess
the supernatural power of controlling it? This is not a physically justified
assumption. The moral of the story is that assuming strict causality is
unphysical and restricts the range of theories we allow ourselves to
explore. Possibly, it excludes the correct solution to the problem of
quantum gravity --- for instance, the theory of quantum gravity with
fakeons, which is predictive and just contains a triplet (the graviton, the
\textquotedblleft Starobinskion\textquotedblright , and the gravifakeon).

These and other facts we have examined expose the notion of causation for
what it truly is: a conceptual mirage, a nonexistent solution to a
nonexistent problem, a camouflage of Plato's \textquotedblleft
ideas\textquotedblright\ (which can act on nature without being part of it).
Not only is the notion of cause incompatible with determinism and quantum
theory, but it blatantly transgresses into metaphysics.

That said, we have shown that concepts such as purely virtual particles, or
fakeons, are not as revolutionary as they might first sound, in the sense
that they do not represent a dramatic departure from the previous
understanding. Once causality is properly assessed, it becomes clear that
fakeons do not bring a true conceptual disruption. The existence of a
minimum time resolution $\tau $ (in the presence of external forces) is not
difficult to accept, and it is consistent with the data if $\tau $ is
sufficiently small. Pre(post)dictivity is just delayed. Finally, a form of
nonlocality has always been present in gauge and gravitational theories
without causing undue concern. The additional nonlocality introduced by
fakeons is only a modest extension of that familiar feature, since both
types of nonlocality arise from parent local theories through projection.

\bigskip

\vskip8truept \noindent {\large \textbf{Acknowledgments}}

\vskip 1truept

We are grateful to U. Aglietti, F. Briscese, L. Buoninfante, L. Modesto, and
the participants of the conference \textquotedblleft \textit{From Puzzles to
New Insights in Fundamental Physics}\textquotedblright , Campagna (SA),
Italy, 23-27 June 2025, for insightful discussions on these topics.

\end{document}